\documentclass[twocolumn,english,aps,prb,showpacs,superscriptaddress,amssymb,amsfonts]{revtex4}
\usepackage[T1]{fontenc}
\usepackage[latin9]{inputenc}
\usepackage{amsmath}
\usepackage{graphicx}
\usepackage{amssymb}
\usepackage{color}
\usepackage{enumerate}

\makeatletter
\newcommand{\be}{\begin{equation}}
\newcommand{\ee}{\end{equation}}                  
\newcommand{\bea}{\begin{eqnarray}}
\newcommand{\eea}{\end{eqnarray}}
\newcommand{\beas}{\begin{eqnarray*}}
\newcommand{\eeas}{\end{eqnarray*}}

\makeatother

\usepackage{babel}

\makeatother

\usepackage{babel}

\begin{document}
 
 \title{Chiral Symmetry Breaking, Deconfinement and Entanglement Monotonicity}

\author{Tarun Grover}

\affiliation{Kavli Institute for Theoretical Physics, University of California, Santa Barbara, CA 93106, USA}

\begin{abstract}
We employ the recent results on the generalization of the $c$-theorem to 2+1-d to derive  non-perturbative results for strongly interacting quantum field theories, including QED-3 and the critical theory corresponding to certain quantum phase transitions in condensed matter systems. In particular, by demanding that the universal constant part of the entanglement entropy decreases along the renormalization group flow (``F-theorem''), we find   bounds on the number of flavors of fermions required for the stability of QED-3 against chiral symmetry breaking and confinement. In this context, the exact results known for the entanglement of superconformal field theories  turn out to be quite useful. Furthermore, the universal number corresponding to the ratio of the entanglement entropy of a free Dirac fermion to that of free scalar plays an interesting role in the bounds derived. Using similar ideas, we also derive strong constraints on the nature of quantum critical points in condensed matter systems with ``topological order''. 
\end{abstract}

\maketitle
\tableofcontents

\section{Introduction}

Strongly interacting field-theories in 2+1-d often arise as a low-energy description of condensed matter systems and also serve as a test-bed for phenomena in 3+1-d. Generally speaking, there are only a few exact results for non-supersymmetric strongly correlated systems in greater than two space-time dimensions. Thus, it is natural to ask whether there exist non-perturbative principles that may not provide solution to particular problems but still help constrain the set of possibilities for a generic problem? An example is the central charge theorem \cite{zamol1986} for two dimensional conformal field theories and its four dimensional version, the $a$-theorem \cite{cardy88, osborn, komar2011}, that strongly constrain the renormalization group flow of conformal field theories. In this paper,  we employ the recent results on the generalization of the central charge theorem to 2+1-d  \cite{myers2011, casini1, casini2, pufu2011, pufuandothers, casini_connect, sachdev}, supplemented with the work of Vafa and Witten \cite{vafa1984}, to derive the low-energy behavior of several strongly interacting 2+1-d theories. In particular, we put an upper bound on the critical number of fermions below which the chiral symmetry breaking occurs in a non-compact QED-3, and an upper bound on the number of fermion flavors required to deconfine fermions in compact QED-3. We also briefly mention generalization to non-abelian gauge theories. Apart from being of general field-theoretic interest, these results have important implications for the stability of several condensed matter systems that go by the name of ``algebraic quantum spin-liquids'' \cite{asl, hermele2004, wenbook, assaad, ryu, cenke, mila2012}. We also derive strong constraints imposed by the entanglement monotonicity on the nature of quantum and classical phase transitions \cite{SachdevBook}  in condensed-matter systems with emergent gauge fields \cite{chen1993, chubukov1994, senthil2004, balents2005, swingle2012, swinglenew}.
  
Let us recall  the central-charge theorem \cite{zamol1986}, or, the ``$c$-theorem'', that has been quite fruitful in detailing the phase diagrams of several 1+1-d systems that have conformally invariant fixed points. The $c$-theorem states that the central charge $c$ of a conformal field theory (CFT) decreases along the renormalization group flow, as one goes from a UV fixed point to an IR fixed point. As an example, a relevant  perturbation to the tricritical Ising point ($c= 7/10 $) can take it to only two possible unitary CFTs: the Ising critical point ($c = 1/2$) or a fully gapped system ($c=0$). There is a similar theorem for four dimensional CFTs, namely Cardy's $a$-theorem \cite{cardy88, osborn} that has been put on rigorous footing  \cite{komar2011} in the recent past.

Recently, there has been progress in developing an analog of $c$-theorem for three space-time dimensions \cite{myers2011, casini1, casini2, pufu2011, pufuandothers, sachdev, casini_connect}. Specifically, Casini and Huerta \cite{casini2} have shown that for Lorentz invariant theories, if one writes the entanglement entropy for a circular region of radius $R$ as $S(R) = \alpha R - \gamma$, the universal constant $\gamma$ decreases along the RG flow.  Specifically, $\gamma_{UV} > \gamma_{IR}$ where UV/IR denote relativistic ultraviolet/infrared fixed points of the RG. A second development has been the conjecture, and a perturbative proof that the universal part $``F"$  of the free energy of a CFT  on a three-sphere $S^3$ decreases  along the RG flow and takes a stationary value $F_0$ at the fixed points \cite{pufu2011, pufuandothers}. Ref.\cite{casini_connect} showed that for conformally invariant systems, $\gamma = F_0$ at the fixed points. Therefore, $\gamma_{UV} > \gamma_{IR}$ already implies that $F_{UV} > F_{IR}$. Below, to derive our results, we will use the entanglement entropy formulation of Refs.\cite{myers2011, casini1, casini2} when thinking about 2+1-d quantum systems, and the $F$-theorem formulation of Refs. \cite{pufu2011, pufuandothers} when working with classical field theories in 3+0 dimensions, though they are equivalent for our purposes, since we will be dealing only with conformally invariant fixed points.

\section{Consequences of Entanglement Monotonicity for QED-3}

\subsection{Critical Number of Flavors for Chiral Symmetry Breaking in Non-Compact QED-3}
Dynamical symmetry breaking is a phenomenon shared by many strongly coupled field theories. In this section, we consider spontaneous generation of chiral mass in non-compact QED-3 \cite{pikarski1984, vafa1984}. The theory is given by:

\be
\mathcal{L}_{QED-3} = \sum_{a = 1}^{N_f} \overline{\psi}_a \left[ -i \gamma_\mu \left( \partial_\mu + i a_\mu \right) \right] \psi_a 
+ \frac{1}{2e^2}F_{\mu \nu} F_{\mu \nu} \label{qed3}
\ee

where $\psi$ is a \textit{four-component} Dirac fermion and $a_\mu$ is a non-compact $U(1)$ gauge field. The four-component notation helps define chiral symmetry in 2+1-d. Organizing $\psi$ as two two-component spinors, $\psi = [\psi_\uparrow\,\, \psi_\downarrow]^T$, and choosing $\gamma_0 = \sigma_3 \otimes \sigma_3, \gamma_1 = i \sigma_3 \otimes \sigma_1, \gamma_2 =  i \sigma_3 \otimes \sigma_2$, the chiral mass is given by $\Delta \mathcal{L} = m \overline{\psi} \psi = \psi^{\dagger}_\uparrow \sigma_3 \psi_\uparrow -  \psi^{\dagger}_\downarrow \sigma_3 \psi_\downarrow$. Such  a mass breaks the global $U(2N_f)$ symmetry of the above action  down to $SU(N_f) \times SU(N_f) \times U(1) \times U(1)$ while preserving the parity and time-reversal. Thus, when generated spontaneously, it leads to $2N_f^2$ number of Goldstone modes. Following Vafa and Witten \cite{vafa1984}, this is the most likely form of spontaneous symmetry breaking pattern for QED-3, an expectation that is also supported by various other analyses \cite{pikarski1984}.

The Lagrangian in Eqn.\ref{qed3} undergoes dynamical chiral symmetry breaking (CSB) only for $N_f < N_{fc,CSB}$ where the precise value of $N_{fc,CSB}$ has been a point of vast discussion \cite{pikarski1984, recentqed, bashir2008}. As we now argue, monotonicity of the entanglement entropy between RG fixed points implies that there indeed exists a finite  $N_{fc,CSB}$ above which CSB cannot occur and furthermore, one can obtain an upper bound  on $N_{fc,CSB}$. 

CSB in a non-compact QED-3 can be thought of as a weak-coupling instability of  a UV fixed  
point consisting of $N_f$  four-component free Dirac fermions and a free photon, to an IR fixed point consisting of $2 N_f^2$ Goldstone modes and a free photon. The universal part of the entanglement entropy for a free Dirac fermion equals $\gamma_{Dirac} = 0.219$ \cite{pufu2011, dowker}. The assignment of $\gamma$ to the free photons as well as the Goldstone modes require a bit more thought since a free photon in three-dimensions is not conformally invariant \cite{sheer}. Indeed, when put on a sphere $S^3$ of size $R$, this lack of conformal invariance leads to a logarithmic dependence of its free energy \cite{sachdev} on $R$ . Since a Goldstone mode is dual to a free photon in three dimensions \cite{zee}, one again expects logarithmic corrections to its entanglement entropy \cite{max}.  This implies that formally, $\gamma = \infty$ for both the Goldstone mode as well as the photon. 

As discussed in Ref.\cite{max}, the presence of subleading logarithmic divergence in the entanglement entropy of Goldstone mode is related to the presence of zero mode for the free scalar theory with Lagrangian $\mathcal{L} = \int \,d^3x\,(\partial_\mu \phi)^2$. There is a physical way to remedy this divergence, which is  to assign an infinitesimal mass $\delta m$ to the Goldstone mode, such that $\delta m \ll l^{-1}$ where $l$ is the size of the region for which the entanglement is being calculated. This assignment of mass $\delta m$ has a direct physical meaning: the phenomena of symmetry breaking is best understood by applying a vanishing small magnetic field that couples to the order-parameter field, and only then taking the thermodynamic limit. The mass $\delta m$ plays the role of this infinitesimal magnetic field in our discussion, and it regularizes the zero mode associated with a free scalar. The main implication of this small mass is that it effectively converts a Goldstone mode to a non-compact conformal scalar, which has a $\gamma$-value of $\gamma_{scalar} = 0.0638$ (Ref.\cite{dowker}) as $l\delta m \rightarrow 0$. 

One might expect that one can similarly assign a small mass to the photon, which via the standard duality \cite{zee}, would  now correspond to a small monopole fugacity. However, there is one important subtlety. The UV fixed point with free photons and fermions has  two potentially relevant couplings: the charge $e$, and the monopole fugacity $\delta m$, which spoils the exact correspondence with the Goldstone mode problem. Furthermore, when $N > N_{fc,CSB}$ (i.e., when interacting QED-3 is a stable fixed point of RG), $\gamma_{IR} = 2N_f \gamma_{Dirac} + \frac{1}{2}\log(\pi N_f/4) + O(1/N_f)$ \cite{sachdev}.  This implies, that at least when $N_f > N_{fc,CSB}$, the assignment $\gamma_{photon} = \gamma_{scalar}$ at the UV fixed point cannot be correct since it will violate the F-theorem. None of these arguments, however, seem to contradict such an assignment for $N_f < N_{fc,CSB}$ (when the free theory flows to the Goldstone phase rather than the interacting QED-3), and below, we will first consider the consequences of such an assignment. This will also provide the spirit of the arguments to come later. Following this, we will present a different argument, which avoids perturbing around the free fixed point and instead employs the flow from \textit{supersymmetric} (SUSY) QED-3 to non-SUSY QED-3.

Let us therefore consider the Lagrangian in Eqn.\ref{qed3} for $N_f < N_{fc,CSB}$. Tentatively assigning $\gamma_{photon} = \gamma_{scalar}$ in the UV, the total value of $\gamma$ at the UV fixed point is given by  \cite{footnote_dirac}

\bea
\gamma_{UV} & = & 2N_f \gamma_{Dirac} + \gamma_{scalar}
\eea

While in the CSB phase, it is given by \cite{footnote_photon}
 
\bea
 \gamma_{IR} & = & 2N_f^2 \gamma_{scalar} + \gamma_{scalar}
 \eea
  
Since  $\gamma_{UV} > \gamma_{IR}$ (Ref. \cite{myers2011, casini1, casini2})
\bea 
N_{fc,CSB} & < & \frac{\gamma_{Dirac}}{\gamma_{scalar}} \nonumber \\
& =  & 3.3056 \label{eq:upperbound}
\eea  

 We note that our argument above is similar in spirit to those presented  by Appelquist et al \cite{appelquist} where it was suggested that the thermodynamic free energy density may serve as an analog of the central charge in three dimensions. However, there are known counter-examples to the monotonicity of free energy density \cite{sachdev_free, footnote_free, pufu2011} and therefore it cannot be used for obtaining such bounds.

\textbf{Bounds via deformation of SUSY QED-3 to obtain QED-3:}

Since the above argument requires dealing with free gauge field in the UV, which formally has an infinite $\gamma$, we now construct a different, more robust argument which instead consists of deforming a superconformal theory to obtain interacting non-SUSY (i.e. conventional) QED-3 in the IR.  The advantage of this approach is that the $\gamma$ for the corresponding supersymmetric theory is finite and it can be calculated \textit{exactly} \cite{kapustin2010, hama2011, jafferis,pufu2011,sachdev}, thus yielding a rigorous upper bound on $N_{fc,CSB}$, under certain reasonable assumptions discussed below.

The supersymmetric theory we consider \cite{ack_klebanov} is  maximally chiral $\mathcal{N}=2$ superconformal QED-3, whose Lagrangian is given by:

\bea
 \mathcal{L}_{\mathrm{SQED-3}} & = & (\partial_\mu a_\nu - \partial_\nu a_\mu)^2 - \frac{i}{4} \overline{\lambda} \gamma_\mu \partial_\mu \lambda  \nonumber   \\ & & + |(i\partial_\mu -ea_\mu) \phi|^2 + \overline{\psi}(-i \partial_\mu - ea_\mu) \gamma_\mu \psi   \nonumber \\
&  & + ie(\phi \overline{\psi} \overline{\lambda} - \phi^* \psi \lambda)  + \frac{D^2}{2} + \frac{eD}{2}|\phi|^2 \nonumber \\
& & + (\partial_\mu \theta)^2 -  e\theta (\overline{\psi} \psi) + e^2\theta^2 |\phi|^2
\eea

\noindent where $\vec{a}$ is the gauge field, $\lambda$ is the (fermionic) gaugino,  $\phi$ is a component complex boson with  $2N_f$ flavors,  $\psi$ is a two-component Dirac fermion with $2N_f$ flavors, $\theta$ is a real scalar and $D$ is an auxiliary field (which can be integrated out to generate quartic terms for the scalar).

Let us now deform the above theory by adding gauge-invariant mass  for $\phi$ ($\propto |\phi|^2) $, as well as a mass for $\theta$ ($\propto \theta^2$). Note that these mass terms retain all the symmetries, in particular, the $SU(N_f)$ flavor symmetry as well as the time-reversal symmetry. This ensures that no explicit mass is generated for the fields $\psi$ and $\lambda$. We can now  integrate out $\theta$ and $\phi$, which generates new interactions for the left over fields $\lambda, \vec{a}, \psi$:

\be 
\mathcal{L} = \mathcal{L}_{QED-3}  - \frac{i}{4} \overline{\lambda} \gamma_\mu \partial_\mu \lambda + \Delta \mathcal{L}_1
\ee

where 

\be 
\mathcal{L}_{QED-3} = \frac{1}{4e^2} (\partial_\mu a_\nu - \partial_\nu a_\mu)^2 + \sum_{\alpha =1}^{N_f}\overline{\psi}_\alpha(-i \partial_\mu - a_\mu) \psi_\alpha 
\ee

and

\be 
\Delta \mathcal{L}_1 \propto \overline{\lambda} \lambda \overline{\psi} \psi 
\ee

$\mathcal{L}_{QED-3}$ has the form of (non-SUSY) QED-3. The terms such as $(\overline{\psi}\psi)^2$, which are generated upon integrating out $\theta$ are already assumed to be part of  $\mathcal{L}_{QED-3}$, since they are allowed by symmetry. For $N_f \gg 1 $ (and bigger than $ N_{fc}$), we know of only one fixed point of this theory, which is the interacting QED-3 fixed point. Therefore, we expect that the charge $e$ will renormalize as one flows  away from the SQED-3 fixed point, such that it attains the value appropriate for (non-SUSY) QED-3 conformal fixed point at the end of RG flow. One still needs to show that  $\Delta \mathcal{L}_1$ is \textit{irrelevant} at the QED-3 fixed point, which we now argue is indeed a very reasonable expectation. This expectation is based on the fact that the scaling dimension of the operator $\overline{\psi}\psi$, in a large-$N_f$ expansion,  is given by $\Delta_{\overline{\psi}\psi} = 2 + \frac{128}{6\pi^2 N_f}$ (Ref. \cite{hermele}). The main thing to notice here is not the exact value of  $\Delta_{\overline{\psi}\psi} $ but the fact that, at the very least, $\Delta_{\overline{\psi}\psi} > 2$ for  $N_f \gg 1$. We assume that this trend holds for the values of $N_f$ we encounter below which are $2N_f \approx 10$. Note that this is rather special to the operator $\overline{\psi}\psi$. In fact, all non-singlet operators of the form $\overline{\psi} T^a \psi$, where $T^a$ is an $SU(2N_f)$ generator acting in the flavor space, have scaling dimensions that are \textit{less} than two in the large-$N_f$ expansion at $O(1/N_f)$ \cite{hermele}. This is also consistent with Vafa-Witten theorem: a time-reversal breaking term of the from $\overline{\psi}\psi$ ought to be less relevant than the terms that possibly  retain the time-reversal.

Assuming $\Delta_{\overline{\psi}\psi} > 2$ for $N_f$ close to $N_{fc,CSB}$, one finds that $\Delta \mathcal{L}_1$ is indeed \textrm{irrelevant} at the QED-3 fixed point (note that $\lambda$ remains a free Dirac fermion and has a scaling dimension of unity). Therefore, one is left with QED-3 and a free Dirac spinor $\lambda$ in the IR. 

Entanglement monotonicity implies,

\be 
\gamma_{SQED-3} > \gamma_{QED-3} + \gamma_{Dirac}  \label{bound1}
\ee

On the other hand,  as discussed earlier, below  $N_{fc,CSB}$, the RG flow from a free UV theory to the QED-3, doesn't stop at the QED-3 fixed point, but instead ends up at the Goldstone mode fixed point, which corresponds to CSB of the form $U(2N_f) \rightarrow U(N_f) \times U(N_f)$. Assuming that the RG flow deforms continuously as $N_f$ is varied,  entanglement monotonicity implies \cite{footnote_photon}, 

\be 
\gamma_{QED-3} > 2N_f^2 \gamma_{scalar} + \gamma_{scalar} \label{bound2}
\ee

Combining Eqns.\ref{bound1} and \ref{bound2}, spontaneous symmetry breaking is allowed \textit{only if},

\be 
(2N_f^2+1) \gamma_{scalar} < \gamma_{SQED-3} - \gamma_{Dirac} \label{bound3}
\ee

$\gamma_{SQED-3,Chiral}$ has been calculated  by Klebanov et al in Ref.\cite{sachdev}. For obtaining the bounds, we can just work with the large-$N_f$ expanded version of $F_{SQED-3}$, since the first three terms yields essentially exact result \cite{sachdev}, which is sufficient for our purposes:

\bea
\gamma_{SQED-3} & = & N_f \log(2)  + \frac{1}{2}\log\left( \frac{N_f\pi}{2}\right)  \nonumber \\ & & + \left( \frac{-1}{4} + \frac{10}{3\pi^2}\right) \frac{1}{N_f}  + O(N_f^{-2})  \label{Fsqed3}
\eea

Using Eqns.\ref{bound3} and \ref{Fsqed3}, one finds that the Goldstone modes cannot be generated when $2N_f >  
13$. Therefore, the above argument implies that $N_{fc,CSB} <7$.

We would like to end this section with two comments. All of our arguments assume that the RG flows evolve smoothly as $N_f$ is varied, even though $N_f$ is not a continuous parameter. There is no apriori justification for this statement. Secondly,  it will be worthwhile to study the flow from SQED-3 to QED-3 in more detail, for example, within a $1/N_f$ expansion, or numerically, to substantiate the RG flow outlined above.

\subsection{Critical Number of Flavors for Deconfinement in Compact QED-3}
The Lagrangian for compact QED-3 is same as Eqn.\ref{qed3}, the only difference being that the instantons in the gauge field configuration are now allowed. Following Polyakov \cite{polyakov1}, in the absence of any matter field, compact QED-3 confines. However, one expects that as the compact gauge field is coupled to fermionic matter fields, the theory deconfines above certain critical number $N_{fc, Dec}$ of flavors of the fermions  \cite{hermele2004} where the precise value of  $N_{fc,Dec}$ required for deconfinement is not known. Here we obtain an upper bound on $N_{fc}$ by employing ideas similar to that in the previous section.

So let us start with compact QED-3 in the UV with massless photons and $N_f$ massless fermions. One can imagine four distinct possibilities for the fate of this theory in the IR:

\begin{enumerate}[I]
\item The theory confines with a mass gap to all excitations.
\item The theory confines while breaking the flavor symmetry $U(2N_f)$ down to some smaller subgroup resulting in massless Goldstone bosons.
\item The theory deconfines with massless fermions in the IR \cite{footnote_deconfine}.
\item The theory deconfines while maintaining gap to the fermions and gauge fields in the IR. \cite{footnote_deconfine}.
\end{enumerate}

In a remarkable paper, Vafa and Witten \cite{vafa1984} argued that in 2+1-d, whenever there exist $N_f>3$ massless fermions coupled to massless gauge bosons in the UV, then there necessarily exist massless particles in the IR. However, the Vafa-Witten theorem does not tell what these massless particles correspond to. They may well be Goldstone modes, or they could also be fermions coupled to the gauge field(s). We now show that entanglement monotonicity can be used to refine Vafa-Witten theorem and obtain a strict bound on $N_f$ for which massless fermions deconfine in the IR (possibility III above). 

To begin with, Vafa-Witten theorem rules out  the possibilities I and IV for $N_f > 3$. Next, assuming that the pattern of flavor symmetry breaking  is \cite{vafa1984, pikarski1984} $U(2N_f) \rightarrow U(N_f) \times U(N_f)$, entanglement monotonicity implies that

\be
\gamma_{QED-3} > 2N_f^2 \gamma_{scalar} \label{bound4}
\ee

Note that in contrast to the discussion in the last subsection, the photon does not survive after symmetry breaking since we are dealing with a compact QED. On the other hand, one can again deform the maximally chiral $\mathcal{N} = 2$ SQED-3 to obtain QED-3 and thus, the Eqn.\ref{bound1} continues to hold. Combining Eqns. \ref{bound1} and \ref{bound4}, one concludes that one can not obtain Goldstone modes in the IR, if $N_f > 6$.  This leave only the possibility III for $N_f > 6$, that is, QED-3 necessarily deconfines when the number of (four-component) flavors exceed six.

\underline{Non-abelian Gauge theories:}
To a good approximation, the bounds derived above all have the form that for $N_f \gtrsim \frac{\gamma_{Dirac}}{\gamma_{scalar}}$, the QED-3 is stable to spontaneous symmetry breaking and/or confinement. The same argument may be generalized to quantum field theories that describe matter coupled to non-abelian gauge fields. Here we only mention an approximate result and leave the derivation of an exact bound (say, by employing the deformation of a SUSY QCD-3), for the future. To the leading order in $N_f$, $\gamma$ for QCD-3 for $2N_f$ flavors of fermions coupled to a non-abelian gauge field with $N_c$ colors is given by $2N_f N_c \gamma_{Dirac}$, while that for the Goldstone phase is given by $2 N^2_f \gamma_{scalar}$. Entanglement monotonicity and Vafa-Witten theorems again imply that  QCD-3 is stable against confinement when $N_f \gtrsim N_c \frac{\gamma_{Dirac}}{ \gamma_{scalar}}$. This is consistent with the general intuition, that as $N_c$ increases, so does the critical number of fermions required for deconfinement \cite{asymptotic, appelquist90}. One can systematically improve this estimate of bound by considering $1/N_f$ corrections to the leading result for the entanglement of non-abelian gauge theories \cite{sachdev}.

\section{Deconfined Vs Conventional Quantum Critical Points}

\begin{figure}
\begin{centering}
\includegraphics[scale=0.34]{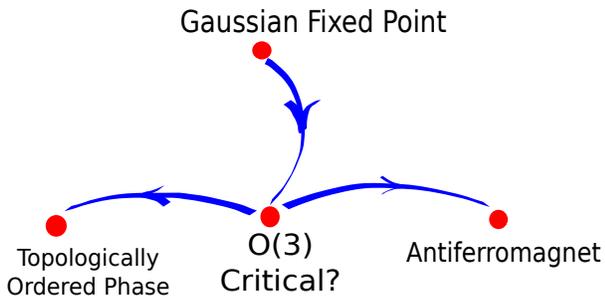}
\par\end{centering}
\caption{ An RG flow that is prohibited due to entanglement monotonicity/F-theorem. The Gaussian fixed point has $\gamma = 3 \times \gamma_{scalar} \approx 0.18$ while the topological ordered phase has $\gamma$ that equals the topological entanglement entropy and is bigger and satisfies $\gamma > \log(\sqrt{2}) \approx 0.35$. This means that the quantum phase transition separating the topological ordered phase and the antiferromagnet cannot be $O(3)$ critical, on general grounds. Similar arguments can be made for several other quantum critical points in condensed matter systems (see the main text).} \label{fig:lattice}
\end{figure}

\textbf{\underline{{Deconfined quantum critical points:}}}
In this section, we will consider a few applications of entanglement monotonicity to interacting spins and/or gauge fields on  lattices in 2+1-d and 3+0-d.

Let us start by posing the following question: for an $SU(2)$ symmetric spin-system in 2+1-d, can there ever be a  quantum phase transition out of a gapped paramagnet that carries one spin-$\frac{1}{2}$ spin per unit cell, such that the transition lies in the conventional $O(3)$ universality class? 

Before attempting to answer the above question, it is perhaps important to understand the backdrop. In  3+0-d classical statistical mechanics, the phase transition between an ordered $O(N)$ magnet and a disordered phase  (i.e. a paramagnet) generically lies in the $O(N)$ universality class. However, quantum mechanical spins carry non-trivial Berry's phase which  may not only lead to a dramatic change in the nature of the magnetic phase transition \cite{readsachdev1989, senthil2004, balents2005}, but perhaps even more strikingly, they  can sometimes disallow the existence of a featureless gapped paramagnet altogether \cite{oshikawa, hastings}. In particular, Hastings \cite{hastings} showed that a paramagnet with   an odd number of spin-$\frac{1}{2}$ spins in the unit cell has  a robust ground state degeneracy on the torus. The non-uniqueness of the ground state is related to the fact that the low-energy theory is a topological quantum field theory and the resulting paramagnet is a ``topologically ordered'' state \cite{wenbook}. Thus, the aforementioned question may be reformulated as: for an $SU(2)$ symmetric spin-system in 2+1-d, can there ever be a  quantum phase transition out of a topologically ordered paramagnet, such that the transition lies in the conventional $O(3)$ universality class? 

A topological ordered paramagnet has a non-zero universal entanglement \cite{levin2006, kitaev2006}, $\gamma_{topo}$. This fact, when supplemented with entanglement montonicity puts a strong constraint on the universality class of the aforementioned transition. One can obtain an upper bound on the $\gamma$ value for the conventional $O(3)$ critical point by flowing down to it from the Gaussian fixed point. The Gaussian point consists of three free scalars and therefore,

\be
\gamma_{O(3)} < 3 \times 0.0638
\ee

If the critical point flows to a topologically ordered paramagnet on one side of the phase diagram, then the $\gamma_{critical}$ corresponding to the phase transition point must satisfy:

\be 
\gamma_{critical} > \gamma_{topo}
\ee

The $\gamma_{topo}$, on the other hand, can only take a discrete set of values, since it is related to the quantum numbers of the excitations that lie above the ground state \cite{levin2006, kitaev2006}. The smallest possible value of $\gamma_{topo}$ is attained by the Laughlin $\nu = 1/2$ state with  $\gamma_{topo} = \log(\sqrt{2}) \approx 0.3466$. Clearly, this is greater than the upper bound for $\gamma_{O(3)}$. Therefore, we conclude that \textit{ such a transition is not allowed to lie  in the $O(3)$ universality class , even though the global symmetry is just $SU(2)$}. Indeed, all the known transitions between the ordered phases and the topologically ordered phases have fractional particles even at the transition which leads to a different value of $\gamma_{critical}$ such that the equation $\gamma_{critical} > \gamma_{topo}$ is always satisfied \cite{swingle2012, swinglenew, sachdev} where $\gamma_{topo}$ is the value for the topologically ordered phase. We note that there already exist several realistic models of frustrated magnets \cite{yao2012, guwen2012, messio, meng}, which seemingly exhibit a direct phase transition between a topologically ordered paramagnet and an $SU(2)$ symmetry broken state.

Similar analysis can be done for phase transitions in bosonic systems that have a global $U(1)$ symmetry corresponding to boson number conservation. For example, the phase transition between a superfluid and a $\nu = 1/2$ fractional quantum Hall state of bosons can not be a conventional O(2) transition because $\gamma_{O(2)} < 2 \times 0.0638$ while $\gamma_{\nu = 1/2} = \log(\sqrt 2) \approx 0.3466$. This is indeed consistent with the known theory for this transition \cite{maissam}. On the other hand, the phase transition between an integer quantum Hall state of bosons and a superfluid is allowed to be in the $O(2)$ universality class, since the integer quantum Hall state has $\gamma = 0$. Indeed, the critical theory for such a transition is just  O(2)  \cite{grover}.

\textbf{\underline{Transitions in classical gauge-matter theories:}}

The same argument as above also applies to classical finite temperature transitions in gauge-matter theories in 3+0 dimensions. For example, consider classical $\mathbb{Z}_2$ gauge-matter at finite temperature on a three-dimensional cubic lattice \cite{fradkin}:

\be 
Z = \sum_{\{\sigma\}, \{s\}} e^{-H(\{\sigma\}, \{s\})}
\ee

where 

\be
H(\{\sigma\}, \{s\}) = -J \sum_{<i,j>} s_i \sigma_{ij} s_j - K \sum_{\Box} \prod_{\Box} \sigma
\ee

This phase diagram of this model is well-understood \cite{fradkin}. The easiest way to establish this phase diagram is to first consider the limit $J = 0$ in which case one obtains a pure gauge theory which is separated from the confined phase via a second-order transition. Similarly, in the limit $K = \infty$, the deconfined phase is separated from the Higgs phase by a second-order transition. As shown in Ref.\cite{fradkin}, the Higgs and the confined phase are one and the same.

 The main point that we want to emphasize here is that though both the aforementioned transitions are generally assumed to be in the Ising universality class, in a strict sense, this is not true.  The free energy $F$ on $S^3$ for the $\mathbb{Z}_2$ deconfined phase is $\log(2)$ (Ref. \cite{witten1989}), while the $F$ for a conventional Ising critical point is bounded from above by $F_{scalar} = 0.0638$. Indeed, by performing a quantum-to-classical mapping, and using the arguments of Ref. \cite{swingle2012}, this transition can be argued to consist of critical 2+1-d Ising CFT coexisting with deconfined (quantum) Ising gauge theory. Using the notation of Ref.\cite{senthilfisher}, the transition may thus be called Ising$^{*}$, and it has the universal $F = F_{\textrm{Ising}} + \log(2)$ where $F_{\textrm{Ising}}$  is the value for the conventional 3D Ising critical point. It is interesting to note that all local operators will have scaling dimensions identical to that at the Ising transition and yet the transition is in a different universality class since the $F$ value differs.
 
The distinction between the regular Ising and Ising* may seem `trivial' in that the CFT with the infinite correlation length ($=$ Ising model) and the TQFT with zero correlation length ($=$ Ising gauge theory) essentially decouple at low-energies. However, presence of topological degrees of freedom at such transitions can have dramatic consequences for the nearby phases as we exemplify now. Consider Kitaev's honeycomb model \cite{kitaev}, where in the absence of any magnetic field and uniform couplings one obtains a nodal $\mathbb{Z}_2$ topologically ordered state. At low-energies, such a state shows decoupling between the gapless fermionic degrees of freedom and the gapped Ising gauge theory, very similar to the case of Ising* transition above. Thus, the value of $\gamma$ for this phase is given by \cite{hongyao} $\gamma_{nodal} = \gamma_{gauge} + \gamma_{fermion} \approx \log(2) + 0.22 \approx 0.91$. Now, this phase is known to be unstable  for infinitesimal value of magnetic field in the $[1 1 1]$ direction to a topological ordered phase with  anyons that obey non-abelian statistics \cite{kitaev} which happens to have a  $\gamma = \log(2)$. Such an RG flow from the gapless state to the gapped topological ordered state would be prohibited due to the entanglement monotonicity, if it were not for the presence of gapped $\mathbb{Z}_2$ gauge sector in the gapless phase. This is because, the fermions alone carry $\gamma_{fermion} \approx 0.22$, too little to allow a direct flow to the non-abelian state with $\gamma = \log(2)$.

\section{Summary and Discussion}

In this paper, we employed the recently results on the generalization of the $c$-theorem to odd space-time dimensions \cite{myers2011, casini1, casini2, pufu2011, pufuandothers, casini_connect, sachdev} to show that compact QED-3 is in the chirally symmetric phase when $N_f > 6$ where $N_f$ is the number of four-component fermionic flavors.  We also showed that compact QED-3 is deconfining when $N_f > 6$ on the number of four-component flavors. We also generalized both of these results to QCD-3 for arbitrary gauge groups. Furthermore, again using the monotonicity of entanglement entropy, we showed that the transitions to topologically ordered paramagnets in an $SU(2)$ symmetric spin-system can never lie in an $O(3)$ universality class. We also pointed out that there exist classical phase transitions where correlation functions of all local operators are identical to that of an Ising transition, yet the critical point does not lie in the conventional Ising universality class.

\textbf{Acknowledgments:} I thank L. Balents, N. Iqbal, V. Kumar, M. Metlistski, J. Polchinski, and especially, I. Klebanov,  for helpful discussions; I also thank M. Hermele, T. Senthil and A. Vishwanath for useful feedback on the draft, and J. McGreevy, S. Sachdev and C. Vafa for helpful email exchanges. I thank organizers and participants of the KITP program ``Frustrated Magnetism and Quantum Spin Liquids''. This research was supported in part by the National Science Foundation under Grant No. NSF PHY11-25915.


\begin{thebibliography}{x}
\bibitem{footnote_dirac} As a side remark, we note that the Dirac fermion in the UV is unaffected by infinitesimal $\delta m$ or the large confinement length and continues to contribute universal entanglement $\gamma_{Dirac}$ at the UV fixed point.
\bibitem{footnote_photon} Note that there is no ambiguity in the assignment $\gamma_{photon} = \gamma_{scalar}$ in the IR since the photon in the IR does not couple to massless fermions (which have been gapped out due to CSB), and thus, the duality to Goldstone modes is exact. Alternatively, one can just provide a mass gap to the photon in the IR, by adding a finite monopole term. This does note modify the bounds on $N_{fc}$ (since $N_f$ is an integer and very small changes on the either side of the inequalities such as Eqn.\ref{bound1} do not matter).
\end{thebibliography}
\end{document}